\documentclass[twoside,twocolumn,9pt]{article}
\usepackage{extsizes}
\usepackage[super,sort&compress,comma]{natbib} 
\usepackage[version=3]{mhchem}
\usepackage[left=1.5cm, right=1.5cm, top=1.785cm, bottom=2.0cm]{geometry}
\usepackage{balance}
\usepackage{mathptmx}
\usepackage{sectsty}
\usepackage{graphicx} 
\usepackage{lastpage}
\usepackage[format=plain,justification=justified,singlelinecheck=false,font={stretch=1.125,small,sf},labelfont=bf,labelsep=space]{caption}
\usepackage{placeins}
\usepackage{float}
\usepackage{fancyhdr}
\usepackage{fnpos}
\usepackage[english]{babel}
\addto{\captionsenglish}{%
  
}
\usepackage{array}
\usepackage{droidsans}
\usepackage{charter}
\usepackage[T1]{fontenc}
\usepackage[usenames,dvipsnames]{xcolor}
\usepackage{setspace}
\usepackage[compact]{titlesec}
\usepackage{hyperref}

\usepackage{float} 
\usepackage{stfloats}

\usepackage{textcomp} 

\usepackage{epstopdf}

\usepackage{booktabs} 

\usepackage[T1]{fontenc}
\usepackage{lmodern}

\definecolor{cream}{RGB}{222,217,201}
\begin{document}

\pagestyle{fancy}
\thispagestyle{plain}
\fancypagestyle{plain}{
\renewcommand{\headrulewidth}{0pt}
}

\makeFNbottom
\makeatletter
\renewcommand\LARGE{\@setfontsize\LARGE{15pt}{17}}
\renewcommand\Large{\@setfontsize\Large{12pt}{14}}
\renewcommand\large{\@setfontsize\large{10pt}{12}}
\renewcommand\footnotesize{\@setfontsize\footnotesize{7pt}{10}}
\renewcommand\scriptsize{\@setfontsize\scriptsize{7pt}{7}}
\makeatother

\renewcommand{\thefootnote}{\fnsymbol{footnote}}
\renewcommand\footnoterule{\vspace*{1pt}%
\color{cream}\hrule width 3.5in height 0.4pt \color{black} \vspace*{5pt}} 
\setcounter{secnumdepth}{5}

\makeatletter 
\renewcommand\@biblabel[1]{#1}            
\renewcommand\@makefntext[1]%
{\noindent\makebox[0pt][r]{\@thefnmark\,}#1}
\makeatother 
\renewcommand{\figurename}{\small{Fig.}~}
\sectionfont{\sffamily\Large}
\subsectionfont{\normalsize}
\subsubsectionfont{\bf}
\setstretch{1.125} 
\setlength{\skip\footins}{0.8cm}
\setlength{\footnotesep}{0.25cm}
\setlength{\jot}{10pt}
\titlespacing*{\section}{0pt}{4pt}{4pt}
\titlespacing*{\subsection}{0pt}{15pt}{1pt}

\fancyfoot{}
\fancyfoot[LO,RE]{\vspace{-7.1pt}\includegraphics[height=9pt]{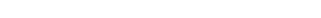}}
\fancyfoot[CO]{\vspace{-7.1pt}\hspace{13.2cm}\includegraphics{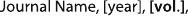}}
\fancyfoot[CE]{\vspace{-7.2pt}\hspace{-14.2cm}\includegraphics{head_foot/RF}}
\fancyfoot[RO]{\footnotesize{\sffamily{1--\pageref{LastPage} ~\textbar  \hspace{2pt}\thepage}}}
\fancyfoot[LE]{\footnotesize{\sffamily{\thepage~\textbar\hspace{3.45cm} 1--\pageref{LastPage}}}}
\fancyhead{}
\renewcommand{\headrulewidth}{0pt} 
\renewcommand{\footrulewidth}{0pt}
\setlength{\arrayrulewidth}{1pt}
\setlength{\columnsep}{6.5mm}
\setlength\bibsep{1pt}

\makeatletter 
\newlength{\figrulesep} 
\setlength{\figrulesep}{0.5\textfloatsep} 

\newcommand{\topfigrule}{\vspace*{-1pt}%
\noindent{\color{cream}\rule[-\figrulesep]{\columnwidth}{1.5pt}} }

\newcommand{\botfigrule}{\vspace*{-2pt}%
\noindent{\color{cream}\rule[\figrulesep]{\columnwidth}{1.5pt}} }

\newcommand{\dblfigrule}{\vspace*{-1pt}%
\noindent{\color{cream}\rule[-\figrulesep]{\textwidth}{1.5pt}} }

\makeatother

\twocolumn[
  \begin{@twocolumnfalse}
{\includegraphics[height=30pt]{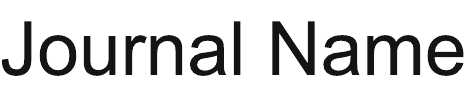}\hfill\raisebox{0pt}[0pt][0pt]{\includegraphics[height=55pt]{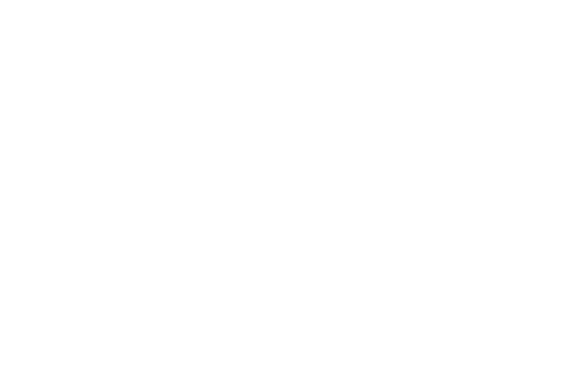}}\\[1ex]
\includegraphics[width=18.5cm]{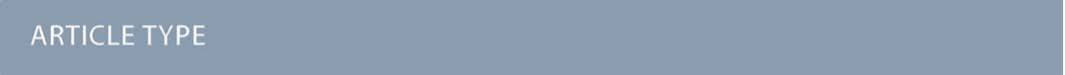}}\par
\vspace{1em}
\sffamily
\begin{tabular}{m{4.5cm} p{13.5cm} }

\includegraphics{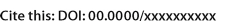} & 
\noindent\LARGE{\textbf{The~chemistry~of~embrittlement:~How~hydrogen\phantom{0.1cm} \mbox{dwindles cohesion in iron during fracture}}}
\\
 & \vspace{0.3cm} \\

 & \noindent\large{Aleksei Egorov$^{\ast}$ 
 } 
 \\

\includegraphics{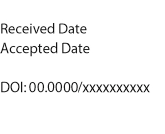} & \\

\end{tabular}

 \end{@twocolumnfalse} \vspace{0.6cm}

  ]

\renewcommand*\rmdefault{bch}\normalfont\upshape
\rmfamily
\section*{}
\vspace{-1cm}


\footnotetext{\textit{$^{\ast}$~Center for Advanced Systems Understanding (CASUS), G\"orlitz, 02826, Germany and Helmholtz-Zentrum Dresden-Rossendorf (HZDR), Dresden, 01328, Germany. \\E-mail: aleksei.egorov.research@gmail.com}}






\sffamily{\textbf{
Normally strong and tough, steel---with iron as its base---falls prey to severe embrittlement when infiltrated by hydrogen; the insights into why are not well understood. This work explores how hydrogen affects bond strength in iron during fracture and why it reduces cohesion between fracture surfaces, laying bare the central role of electrostatic repulsion.
}
}\\


\rmfamily 


In 1875, British metallurgist William Jonson, through a series of experiments, evinced---\textit{it is hydrogen that embrittles iron}\cite{johnson1875ii}. And embrittles severely: a minuscule amount of hydrogen---a few parts per million---can impair iron's ability to withstand fracture, sometimes reducing it by an order of magnitude\cite{dozen_times_more_brittle_LI201815575}. In the century and a half following Jonson's revelation, several explanations were surmised\cite{pfeil1926effect_hede, westlake_hydrides_osti_4173745,Beachem1972, LYNCH1979_aide, BIRNBAUM1994191_help, vacancies_theory_2001590, Song_Curtin_NatMat_2013, gabor_theory_SHISHVAN2020103740}, but the unequivocal mechanism behind hydrogen embrittlement remains unclear\cite{sutton2024elasticitybook}.


Embrittlement implies crack growth. In brittle crystals---like silicon at room temperature\cite{Lawn1980_sharp_Si_crack} or 2D ReS$_2$\cite{2D_crack_tip_PhysRevLett.125.246102}---under load, bonds between atoms at the crack tip break; the tip cleaves and propagates, leaving behind two strips of newly minted surfaces (fracture surfaces). 
One of the proposed mechanisms of embrittlement puts forth that hydrogen weakens iron-iron bonds, rendering them easier to break (or \mbox{\textit{decohere}}), thereby fostering crack tip cleavage and propagation\cite{pfeil1926effect_hede,oriani_hede_1972}.
This hypothesis gained a foothold in experiments on pre-cracked single crystals\cite{VEHOFF1980,VEHOFF1986_2nd,Chen_and_Gerberich_1991_single_crystal_fe}.

\begin{figure}[htb]
\centering
 \includegraphics[width=0.47999999\textwidth]{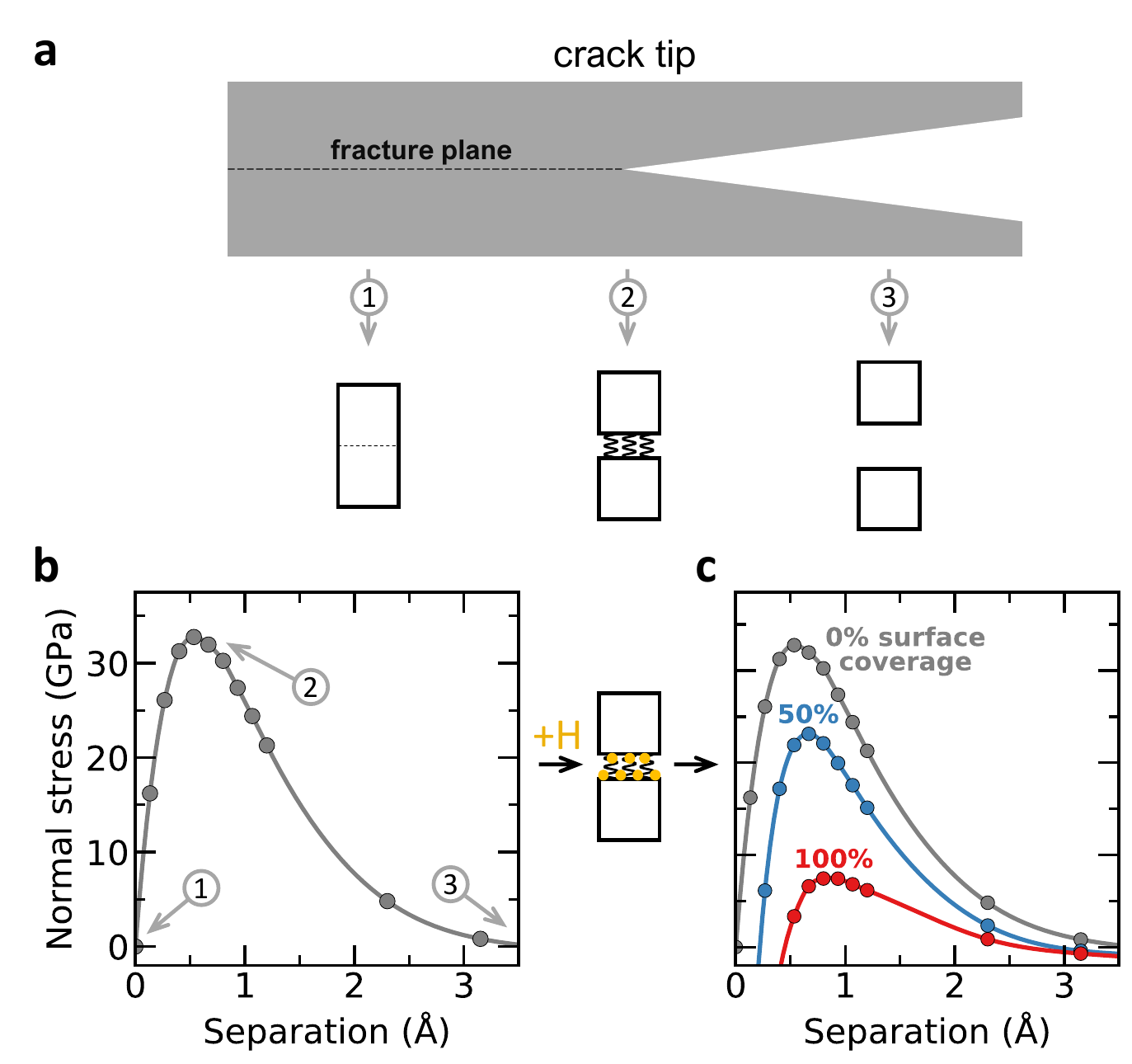}
  \caption{(a) Schematic of the crack tip and the model representing crack propagation as a transformation of the bulk crystal into two parted halves. (b)~Traction-separation (TS) curve for pure iron. (c)~TS~curves for iron with varying hydrogen coverages on the fracture surfaces\cite{note_excess_volume,excess_VANDERVEN20041223,excess_guzman2020hydrogen}. 
  }
  \label{fig:1}
\end{figure}

Direct scrutiny of crack tips in quantum-mechanical simulations, such as Density Functional Theory (DFT)\cite{PhysRev.136.B864-DFT-1,PhysRev.140.A1133DFT-2}, remains unfeasible because such simulations often demand hundreds of thousands of atoms---far beyond the reach of DFT. But fracture mechanics offers an option. \mbox{Since crack propagation} transforms the bulk crystal (ahead of the advancing crack tip) into two parted halves with two newly formed surfaces facing each other (behind the passed crack tip), rather than examining all the atomic intricacies of the crack tip, one might focus on \mbox{\textit{separation}}\cite{Möller2018_ts_explained}. This is brought about by slicing the simulation cell along a specific plane (the fracture plane) and rigidly pulling the two halves apart (Fig.~\ref{fig:1}a). When stress (force per unit area) normal to the fracture plane is computed \textit{en route}, the result is called the traction-separation (TS) curve (Fig.~\ref{fig:1}b). The TS curve imbues the essential physics behind brittle fracture, with its \textit{peak} delineating the maximum load the material can withstand\cite{curtin_ts_curve_explained_2005methods}.
The ease of separation---and thus the ease of crack tip cleavage and propagation---reflected by the height of the TS curve's peak is set by the strength of the iron-iron bonds across the fracture plane\cite{Robb_THOMSON_book_1986}. 
Previous efforts to examine how hydrogen affects bonding in iron focused on bulk crystals, vacancies, grain boundaries, clusters, and free surfaces\cite{bonding_1_Itsumi1996,bonding_2_Itsumi_Ellis_1996,JUAN_hoffman_19991,Juan_GB_COOP_GESARI2002207}, but TS curves have remained unexplored.


\begin{figure}[H]
    \centering
    \includegraphics[width=0.38\textwidth]{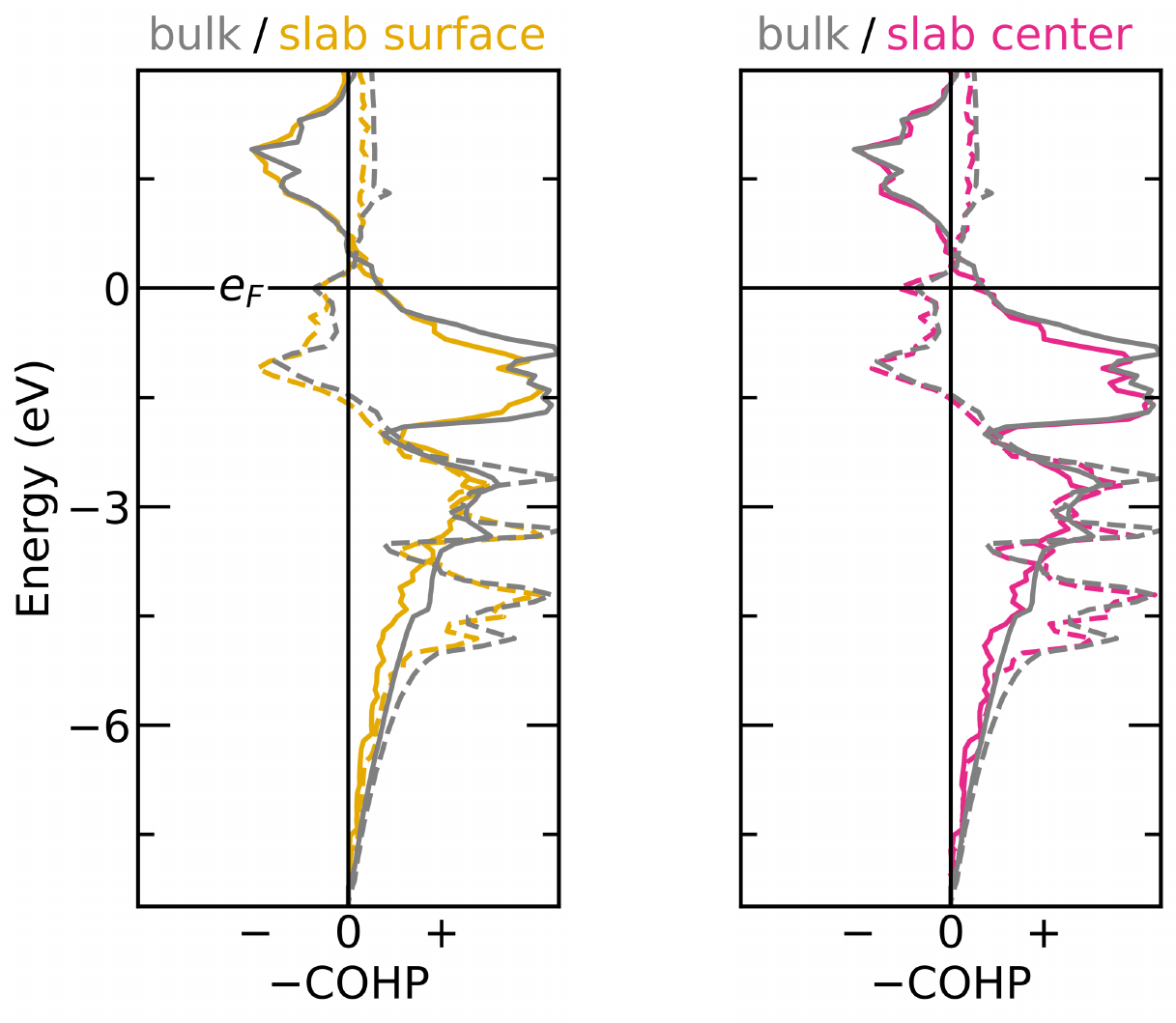}
    \caption{ The spin-resolved COHP for bulk bcc \mbox{Fe and those for} a pair of nearest-neighbor atoms within the center of the ten-layer slab and for a pair with one atom at the surface and the other in the subsurface; majority spin channels are depicted as full lines, minority as dashed lines. In this and subsequent figures, bonding levels are on the right and antibonding on the left\cite{pymatgen_cohp_curves_note, pymatgen_ONG2013314}. The COHP values range from $\mathrel{-}$0.45 to +0.45~eV.}
    \label{fig:slab_conv}
\end{figure}

One of the most robust ways to elucidate the nature of chemical bonding is to exploit the crystal orbital Hamilton population (COHP) analysis\cite{cohp_1993}.
This method translates the delocalized language of DFT plane waves into the localized language of atomic orbitals, enabling one to disentangle bonding (stabilizing) and antibonding (destabilizing) contributions to the electronic band-structure energy\cite{review_muller_drons_D5SC02936H}. By integrating COHP curves, we can further elicit bond strength. Over the past decades, COHP analysis has held its own across a broad swath of applications: revealing the chemical roots of ferromagnetism\cite{drons_ferromag_2nd_paper}, gauging the covalency of hydrogen bonding\cite{Deringer_ChemComm_H_bonds_C4CC04716H}, decoding the origins of metal carbides' ductility\cite{music_dislo_cohp_PhysRevB.75.174102}, and much more\cite{dronskowskibook2023chemical}. And yet, for exploring hydrogen embrittlement of iron, it remains untapped.

I seek to \textit{bring together chemical and fracture mechanics tools} by leveraging COHP analysis to examine bonding at the peak of the representative TS curve for body-centered cubic (bcc) iron (with eight nearest neighbors at 2.454 Å and six at 2.834 Å\cite{note_nn_dist}), with and without hydrogen. This marriage can reveal how the ease of cleavage intertwines with bond strength, shedding light on the roots of embrittlement in pre-cracked single crystals\cite{Vehoff1997}.



I explored the TS curve for the (110) fracture plane, often featured in experiments\cite{Nakasato1978_110_crack_surf}; hydrogen atoms were placed at the threefold surface sites---the most stable sites according to DFT\cite{carter_2003_Acta_JIANG200385,Carter_2004_PhysRevB.70.064102}---and relaxed at zero separation\cite{note_relax_params,FIRE_Bitzek_PhysRevLett.97.170201,note_relax_importance,excess_guzman2020hydrogen}.
DFT calculations were produced using VASP\cite{vasp-1,vasp-2,vasp-3,ase_workflow_note,larsen2017ase} with the projector-augmented-wave (PAW) method\cite{PAW} and the Perdew-Burke-Ernzerhof (PBE)-GGA exchange-correlation functional\cite{PBE}.
To ensure convergence, a plane-wave energy cutoff of 550~eV and a k-point spacing of 0.16~Å$^{-1}$ (0.12 for atomic relaxations) were adopted\cite{smearing,Methfessel_Paxton_PhysRevB.40.3616,tetrahedron_PhysRevB.49.16223}. The LOBSTER package was employed for COHP and Mulliken population analyses\cite{lobster_1, lobster_3,lobster_2,mulliken_C9RA05190B}, and a ten-layer slab was used to obtain the TS curves. To ensure that ten layers are sufficient, I computed the COHP for (i) bulk bcc~Fe, (ii) a pair of nearest-neighbor atoms at the center of the slab, and (iii) a pair with one atom at the surface and the other in the subsurface.
For the surface-subsurface, COHP deviates from the bulk; at the slab center, it largely follows the bulk (Fig.~\ref{fig:slab_conv}). So the slab's center and the bulk behave alike, and ten layers should suffice.






From the computed TS curves (Fig.~\ref{fig:1}c), one can see that hydrogen boosts decohesion by sapping the peak stress required to separate the two halves; the higher the coverage, the lower the peak stress. As separation becomes easier, the crack tip can cleave and propagate more readily. Dissecting the nature of the Fe--Fe chemical bonds that hold the two halves together can help unravel why, when hydrogen is present, the peak stress is attenuated. But first, let us take a look at the bonds involving hydrogen itself.

\begin{figure*}[ht]
\centering
 \includegraphics[width=0.989999\textwidth]{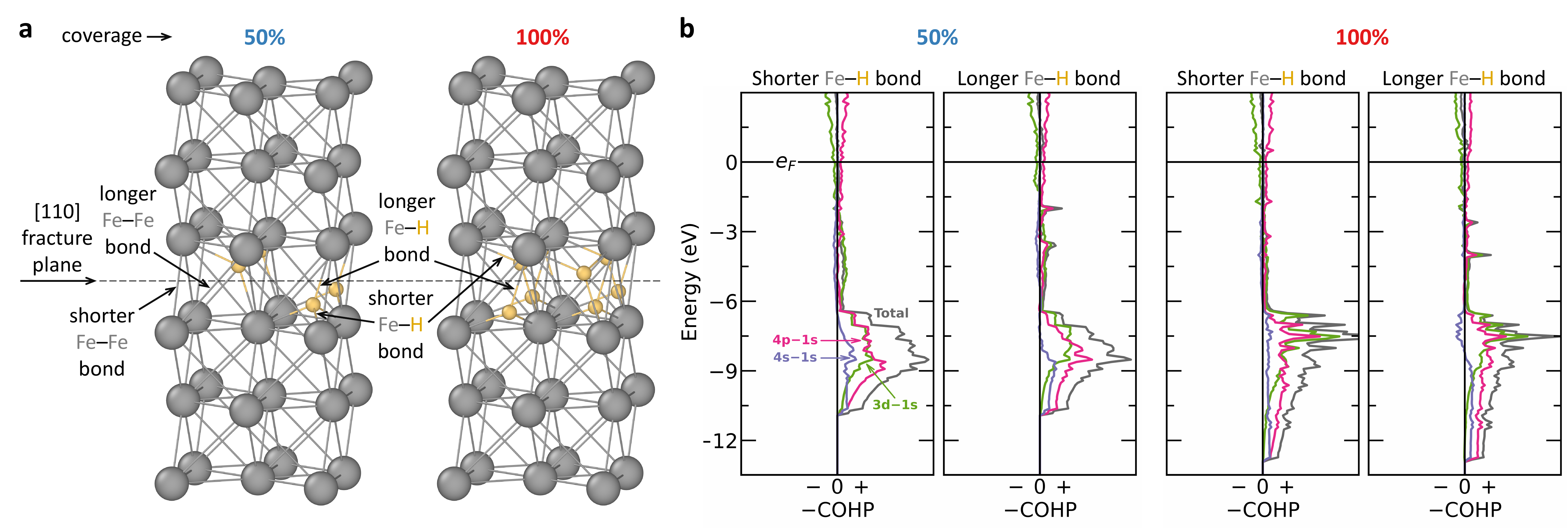}
  \caption{(a) Atomic structures of the fracture planes for computing traction-separation (TS) curves at 50\% and 100\% hydrogen coverage. The separation between the two halves equals the separation at the peak of the pure-iron TS curve depicted in Fig.~\ref{fig:1}. Figure (a) highlights the two distinct types of \mbox{Fe--Fe} bonds connecting the two halves, as well as the two types of Fe--H bonds that form when the two halves meet. (b)~Corresponding COHP plots for the two distinct \mbox{Fe--H} bonds, depicting bonding (+) and antibonding ($\mathrel{-}$) contributions. In addition to the total, orbital-resolved COHPs are displayed for the Fe orbitals that interact with the H 1s orbital. The COHP values range from $\mathrel{-}$0.8 to +0.8~eV. 
  }
  \label{fig:2}
\end{figure*}

Each hydrogen atom at the fracture surface forms three bonds with nearby iron atoms (``shorter Fe--H bond'' in Fig.~\ref{fig:2}a) but also one bond with the nearest iron atom on the opposing surface (``longer \mbox{Fe--H} bond'' in Fig.~\ref{fig:2}a). In both cases, and for both 50\% and 100\% coverage, the COHP plots revealed that bonding interactions dominate below the Fermi level---a hallmark of strong bonds (Fig.~\ref{fig:2}b). All iron orbitals---3d, 4p, and 4s---take part in bonding with hydrogen's 1s orbital, with the first two contributing most; interactions approach an energy of $\mathrel{-}$13.6 eV---the H 1s level. The COHP humps are lower for longer bonds than for shorter ones due to smaller overlap. For 100\% coverage, compared to 50\%, the COHP is more spread out because hydrogen atoms are closer together---this increases overlap and results in wider bands and thus a wider COHP\cite{sung_hoffmann_coverage_1985}.
\mbox{Hydrogen atoms do not} form bonds with each other, as reflected in the negligible COHP between them (not shown). In line with previous studies\cite{Olson_PhysRevB.62.13938,JUAN_hoffman_19991, charge_transfer_S_Cottenier_MEIER2026152840},~the formation of \mbox{Fe--H} bonds entails charge transfer from the surface iron atoms to hydrogen---owing to the latter being more electronegative\cite{Rahm2019Electronegativity}. In terms of Mulliken charges\cite{mulliken_C9RA05190B}, each hydrogen gains 0.25$e^-$ and 0.23$e^-$ for the 50\% and 100\% coverage.

\begin{figure*}[!b]
\centering
 \includegraphics[width=0.989999\textwidth]{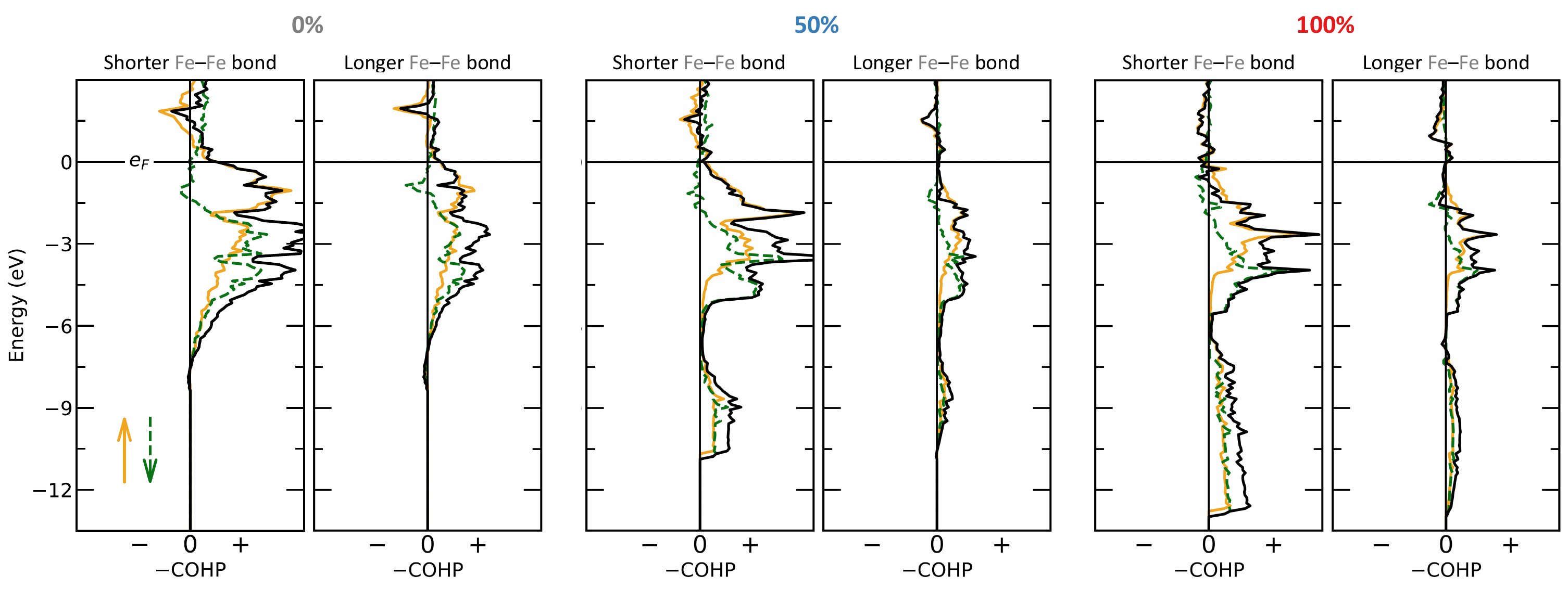}
  \caption{Spin-resolved and total (black) COHP at various hydrogen coverages of the fracture surfaces for the two types of Fe--Fe bonds marked in Fig.~\ref{fig:2}a. \mbox{At 100\%} coverage, the total COHP reveals destabilizing antibonding interactions at the Fermi level. 
  At 50\% coverage, the otherwise equivalent longer Fe--Fe bonds split into two inequivalent types: one adjacent to an H atom and one with H farther away (see Fig.~\ref{fig:2}a). The displayed COHP corresponds to the longer bond adjacent to H (the other bond is largely unaffected by hydrogen, with COHP closely matching that for zero coverage). The COHP values range from $\mathrel{-}$0.45 to +0.45~eV, the same as in Fig.~\ref{fig:slab_conv}.
  }
  \label{fig:3}
\end{figure*}


Now that we have explored the bonds that form between iron and hydrogen across the two fracture surfaces, we arrive at the question of how the formation of these bonds affects the bonding between iron atoms that bridge the surfaces. Because iron is magnetic, with majority and minority spin channels differing markedly and displaying distinct bonding properties\cite{drons_landrum_1999ferromagnetism}, it is imperative to look at spin-resolved COHP (for the Fe--H bonds, the difference between majority and minority spin channels is meager, so spin-resolved COHP was not shown).
For iron without hydrogen (Fig.~\ref{fig:3}-left), the COHP profile resembles that of pure bulk (Fig.~\ref{fig:slab_conv}), except for shrunk-down peaks and troughs due to less overlap between iron orbitals across the fracture plane at the TS curve's peak separation.
For hydrogen-covered surfaces, the first feature to catch the eye is the humps at low energies, arising due to iron orbitals participating in bonding with hydrogen. But the most salient hallmark is a \textit{drastic contraction of the bonding COHP}.
Moreover, the majority-spin bonding levels near the Fermi energy, which, for the clean surface, exceed the antibonding minority-spin levels, begin, as hydrogen coverage increases, to dwindle and, at full coverage, for a shorter Fe--Fe bond, fail to outweigh the antibonding levels. This results in net antibonding interactions at the Fermi level---\textit{a mark of structural instability}\cite{drons_landrum_1999ferromagnetism} (\mbox{von Appen} and colleagues observed a similar kickoff of Fe--Fe bond instability but in bulk fcc Fe--Mn alloys\cite{hickel_Fe-Mn_https://doi.org/10.1002/jcc.23742}).

By integrating the COHP curves, we can elicit a measure of bond strength, the ICOHP. It revealed that when hydrogen is present, the strength of \mbox{Fe--Fe} bonds linking the two halves plummets. Compared to pure iron, it drops by 20\% at half coverage and by 40\% at full. \textit{Hydrogen weakens the \mbox{Fe--Fe} bonds}.

As Roald Hoffmann underlined in his seminal work\cite{Hoffmann_surf_RevModPhys.60.601}, \mbox{``[on surfaces]} metal--adsorbate bonding is accomplished at the expense of bonding within the metal and the adsorbent molecule.'' And, as the current study laid bare, this tenet proved just as valid for the metal--metal bonds that bridge two surfaces.




One might think that weaker Fe--Fe bonds between fracture surfaces would entail easier separation of the two halves. But that is not what is happening.
To separate the two halves, the \mbox{Fe--Fe} bonds across the fracture plane must be broken---so this part becomes easier. But looking at the fracture planes (Fig.~\ref{fig:2}a), we see the two halves are also bridged by Fe--H bonds (``longer Fe--H bond'') and separation would require breaking them too. To get the net effect, we need to sum the strengths of both Fe--Fe and Fe--H bonds. The summation revealed that the strength of Fe--H bonds outweighs the decrease in Fe--Fe bonding. Compared to pure iron, the combined bond strength across the fracture plane rises by 17\% at half-coverage and by 28\% at full.
Oddly enough, \textit{hydrogen strengthened the covalent bonding between fracture surfaces}. Yet it has found another way to compromise their cohesion.

\begin{table}[t]
\small
\caption{
Strength of the Fe--Fe and Fe--H covalent bonds (ICOHP) bridging the two fracture surfaces at the separation corresponding to the peak stress on the traction-separation (TS) curve for pure iron (Fig.~\ref{fig:1}), and the strength of the electrostatic repulsion between the fracture surfaces; the more negative the energies, the greater the stability. All values are given per unit fracture surface area\cite{madelung_energies_values}.
*~indicates whether there is a net (destabilizing) antibonding interaction at the Fermi level for the shorter/longer bonds. (At 50\% coverage, electrostatic repulsion does not fully offset the strengthening of covalent bonds. Nevertheless, the TS curve's peak in Fig.~\ref{fig:1}c is lower than in pure iron because of a volumetric effect: hydrogen atoms take up space and push surrounding iron atoms outward\cite{h_press_outward_Song_Curtin_2010}.)
}
\label{tbl:1}
\begin{tabular*}{0.48\textwidth}{@{\extracolsep{\fill}}lllll}
\toprule
Surface 
& \multicolumn{3}{c}{Bonding strength, ICOHP (eV/\AA$^{2}$)}
& Electrostatic \\
\cmidrule(lr){2-4}
coverage (\%) 
& Fe--Fe    
& Fe--H 
& Total
& repulsion (eV/\AA$^{2}$) \\
\midrule
0   & $-$$0.89^{~-/-}$ & NA & $-$0.89 & $-$0.01 \\
50  & $-$$0.71^{~-/-}$ & $-$0.33 & $-$1.04 & $+$0.08 \\
100 & $-$$0.53^{~*/-}$ & $-$0.61 & $-$1.14 & $+$0.31 \\
\bottomrule
\end{tabular*}
\end{table}


Stepping back to the charge transfer, we see that the hydrogen layers on both fracture surfaces---by drawing electrons out of the iron atoms---become negatively charged. Two negatively charged layers repel---in principle, aiding separation. But there are other electrostatic interactions because the iron surface atoms, giving up electrons, become positively charged. We end up with a positively charged iron layer, covered by a negatively charged hydrogen layer, facing another negatively charged hydrogen layer on the opposite side, backed by another positively charged iron layer. Same-sign layers repel; opposite-sign layers attract. Their interplay is reflected by the Madelung energy\cite{mulliken_C9RA05190B}. To gauge whether there is net electrostatic repulsion, I compared the Madelung energy of the structure at the TS curve with that of the two free surfaces, thereby isolating interactions between ions on the same fracture surface from those with counterparts on the opposing side. The results revealed strong electrostatic repulsion, especially at full coverage\cite{madelung_energies_values} (and fracture surfaces are expected to be fully covered\cite{Vehoff1997,full_h_coverage_Wang2014HAdsorptionFe}). This repulsion counteracts~the enhanced covalent bonding and, at full coverage, outweighs it, tilting \textit{the net effect of hydrogen toward easier separation} (Table~\ref{tbl:1}).


These conclusions resonate with experimental\cite{VEHOFF1980,VEHOFF1986_2nd, Chen_and_Gerberich_1991_single_crystal_fe} and atomistic simulation results\cite{Song_Curtin_NatMat_2013,egorov2025crackcrackhydrogenfavors_other_ref_format} for single crystals. For instance, recent \mbox{large-scale} simulations with a near-DFT-accurate machine-learning-based interatomic potential revealed that (1) hydrogen triggers the breaking of iron-iron bonds at the crack tip (hydrogen flows in, bonds break; it flows out, and bonds reform), and (2) the more hydrogen present, the lower the load required for cleavage\cite{egorov2025crackcrackhydrogenfavors_other_ref_format}. These atomistic simulations allowed us to  \textit{observe how} cleavage unfolds; the current work helps to \textit{understand why}.

Hydrogen embrittlement of iron likely arises from a concerted effort of various phenomena operating at different scales\cite{sutton_2017_discussion_mechanisms}. Here, I corroborated that hydrogen reduces cohesion between fracture surfaces not by sapping the covalent bonds that hold them together, but by inducing electrostatic repulsion between them. 
It is doubtful that reduced cohesion is solely responsible for hydrogen embrittlement\cite{Beachem1978Cleavage,Sutton_DBT_03042026}, yet it is likely a piece of the puzzle.

\section*{Acknowledgements}

I'm grateful to Richard Dronskowski, Rebecca Janisch, and Adrian Sutton for their insightful comments on the manuscript. Calculations were performed on the Vulcan computer cluster at ICAMS, Ruhr-Universität Bochum.




\bibliography{rsc} 
\bibliographystyle{rsc} 

\end{document}